\documentclass[11pt]{article}
\usepackage{amssymb}
\begin{document}

\title{Lie group classifications and exact solutions for time-fractional Burgers equation}
\author{Guo-cheng Wu\footnote{Corresponding
author, E-mail:~wuguocheng2002@yahoo.com.cn. (G.C. Wu)}
\vspace{4mm}\\
College of Textile, Donghua University,  Shanghai 201620, P.R. China;\\
Modern Textile Institute, Donghua University, Shanghai 200051, P.R.
China.}
  \date{}
\maketitle

--------------------------------------------------------------------------------------------------------------------\\
\leftline{\bf\ Abstract}

Lie group method provides an efficient tool to solve nonlinear
partial differential equations. This paper suggests a fractional Lie
group method for fractional partial differential equations. A
time-fractional Burgers equation is used as an example to illustrate
the effectiveness of the Lie group method and some classes of exact
solutions are obtained.

 \vspace{0.5cm} \noindent{\bf PACS}\quad  02.20.Tw; 45.10.Hj

\vspace{0.5cm} \noindent{\bf Key words}\quad  Lie group method;
Fractional Burgers equation; Fractional characteristic method

--------------------------------------------------------------------------------------------------------------------\\
\section{Introduction}

Many methods of mathematical physics have been developed to solve
differential equations, among which Lie group method is an efficient
approach to derive the exact solution of nonlinear partial
differential equations.

Since Sophus Lie's group analysis work more than 100 years ago, Lie
group theory has become more and more pervasive in its influence on
other mathematical disciplines [1, 2]. There are, however, there few
applications of Lie method in fractional calculus. Then a question
may naturally arise: is there a fractional Lie group method for
fractional differential equations?

Some researchers investigated Lie group method for fractional
differential equations in sense of the Caputo derivative and derived
scaling transformation and similarity solutions [3--5]. Considering
the classical Lie group method, method of characteristic is used to
solve symmetry equations.  Recently, with the modified
Riemann-Liouville derivative [6--8], we first propose a more
generalized fractional characteristic method [9] than Jumarie's
Lagrange method [10]. Using our fractional characteristic method,
the generalized symmetry equations generating by the prolongation
technique can be solved, and a fractional Lie group method was
presented for an anomalous diffusion equation [9].

In this study,  we investigate a simplified version of the
fractional Burgers equations [5]

\begin{equation} u_t^{(\alpha )}  =
u_{xx}  + u_x^2 ,~x\in (0,~\infty),\;\;0{\rm{ <
}}t{\rm{,}}\;\;0{\rm{ < }}\alpha <1,
\end{equation}
and derive its group classifications. In order to investigate the
local behaviors of the above equation, the fractional derivative is
in the sense of the modified Riemann-Liouville [6--8].

\section{Fractional Calculus and Some Properties} From the viewpoint of Brown motion, Jumarie proposed  the modified Riemann-Liouville derivative [6--8],

\begin{equation} D_{x }^\alpha
f(x) = \frac{1}{{\Gamma (n - \alpha )}}\frac{{d^n }}{{dx^n
}}\int_{\,0 }^{\,x}  (x - \xi )^{n - \alpha-1 } (f(\xi ) - f(0
))\;d\xi , ~~n-1{\rm{ < }}\alpha <n, \label{eq3}
%(3)
\end{equation}
where the derivative on the right-hand side is the Riemann-Liouville
fractional derivative and $n\in Z^{+}$.

(a) Fractional Taylor series

Recently, Jumarie-Taylor series [11] was proposed
\begin{equation}
df(x) = \sum\limits_{i = 1}^\infty  {\frac{{h^{k\alpha }
}}{{(k\alpha )!}}f^{(k\alpha )} } (x),~~0{\rm{ < }}\alpha <1.
\end{equation}
Here $f(x)$ is a $k\alpha$-differentiable function and $k$ is an
arbitrary positive integer.

Taking $k=1$, $f(x)$ is a $\alpha$-differentiable function. We can
derive that

\begin{equation}df(x)=\frac{D_x^\alpha f(x)(dx)^{\alpha}}{\Gamma (1 + \alpha
)}.
\end{equation}

(b) Fractional Leibniz product law

If we set $D_x^\alpha  u(x)$ and $D_x^\alpha  v(x)$ exist, we can
readily find that

\begin{equation}
D_x^\alpha (uv) = u^{(\alpha )} v + uv^{(\alpha )} {\rm{.}}
\end{equation}

The properties of Jumarie's derivative were summarized in [11]. The
extension of Jumarie's fractional derivative and integral to
variations approach by Almeida et al. [12, 13]. Fractional
variational interactional method and Adomian decomposition method
are proposed for fractional differential equations [14, 15].

(c) Integration with respect to $(dx)^\alpha $

\begin{equation}
_0 I_x^\alpha  f(x) = \frac{1}{{\Gamma (\alpha )}}\int_0^x  (x - \xi
)^{\alpha  - 1} f(\xi )d\xi  = \frac{1}{{\Gamma (\alpha  +
1)}}\int_0^x
 f(\xi )(d\xi )^\alpha  ,0 < \alpha  \le 1.
\label{eq5}
%(5)
\end{equation}

 (d) Generalized Newton-Leibniz Law

Assume $D^{\alpha}_{x}f(x)$ is an integrable function in the
interval $[0, a]$. Obviously,

\begin{equation}
\frac{1}{\Gamma(1+\alpha)}\int^{a}_{0}D^{\alpha}_{x}f(x)(dx)^{\alpha}=f(a)-f(0),
0 < \alpha  < 1,
\end{equation}

\begin{equation}
\frac{1}{\Gamma(1+\alpha)}\int^{x}_{0}D^{\alpha}_{\xi}f(\xi)(d\xi)^{\alpha}=f(x)-f(a),
\end{equation}
and
\begin{equation}
\frac{D^{\alpha}_{x}}{\Gamma(1+\alpha)}\int^{x}_{0}f(\xi)(d\xi)^{\alpha}=f(x),~~0
< \alpha  < 1.
\end{equation}

(e) Some other useful properties

\begin{equation}
f^{(\alpha )}([x(t)])  = \frac{{df}}{{dx}}x^{(\alpha )} (t),~~ 0 <
\alpha  < 1,
\end{equation}
\begin{equation}
 D_x^{\alpha } x^\beta   = \frac{{\Gamma (1 +
\beta )}}{{\Gamma (1 + \beta  - \alpha )}}x^{\beta  - \alpha } ,~~0
< \beta < 1,
\end{equation}
\begin{equation}
\int {(dx)^{^\beta  } }  = x^{^\beta  }. \end{equation}

The above properties (a)--(d) can be found in Ref. [11]. We must
point out that $f(x)$ should be differentiable w.r.t $x$ in Eq.
(10), and $x^{^\beta  }$ is an $\alpha$ order function in Eq. (11).

\section{A Characteristic Method for Fractional Differential Equations}

It is well known that the method of characteristics has played a
very important role in mathematical physics. Preciously, the method
of characteristics is used to solve the initial value problem for
general first order. With the modified Riemann-Liouville derivative,
Jumaire ever gave a Lagrange characteristic method [10], in which
the time-fractional order equals to the space-fractional order. We
present a more generalized fractional method of characteristics and
use it to solve linear fractional partial equations.

Consider the following first order equation,
\begin{equation}
a(x,t)\frac{{\partial u(x,t)}}{{\partial x}} + b(x,t)\frac{{\partial
u(x,t)}}{{\partial t}} = c(x,t). \label{eq7}
%(7)
\end{equation}

The goal of the method of characteristics is to change coordinates
from ${\rm{(}}x,\;t{\rm{)}}$ to a new coordinate system ${\rm{(}}x_0
,\;s{\rm{)}}$ in which the partial differential equation becomes an
ordinary differential equation along certain curves in the $x - t$
plane. The curves are called the characteristic curves. More
generally, we consider to extend this method to linear space-time
fractional differential equations
\begin{equation}
a(x,t)\frac{{\partial ^{^\beta  } u(x,t)}}{{\partial x^\beta  }} +
b(x,t)\frac{{\partial ^\alpha  u(x,t)}}{{\partial t^\alpha  }} =
c(x,t),~~0 < \alpha ,\beta < 1. \label{eq8}
%(8)
\end{equation}

With the fractional Taylor's series in two variables [11]

\begin{equation}
du = \frac{{\partial ^{^\beta  } u(x,t)}}{{\Gamma (1 + \beta
)\partial x^\beta  }}(dx)^{^\beta  }  + \frac{{\partial ^\alpha
u(x,t)}}{{\Gamma (1 + \alpha )\partial t^\alpha  }}(dt)^\alpha
,\;\;0 < \alpha ,\;\beta <1, \label{eq9}
%(9)
\end{equation}
similarly, we derive the generalized characteristic curves
\begin{equation}
\frac{{du}}{{ds}} = c(x,t),
\end{equation}
 \begin{equation}
\frac{{(dx)^{^\beta  } }}{{\Gamma (1 + \beta )ds}} =
a(x,t),\label{eq10}
%(10)
\end{equation}
\begin{equation}
\frac{{(dt)^\alpha  }}{{\Gamma (1 + \alpha )ds}} = b(x,t). \\
\\
\end{equation}
Eqs. (16)--(18) can be reduced as Jumaire's Lagrange method of
characteristic if $\alpha  = \beta $ in [10].

\section{A Fractional Lie Group Method }
 In the classical Lie method for partial differential equations, the one-parameter Lie group of transformations in
\begin{math}{\rm{(}}x,\;t,\;u)\end{math} is given by
\begin{displaymath}
\begin{array}{l}
\tilde x = x + \varepsilon \xi (x,t,u) + O(\varepsilon ^2 ), \\
\tilde t = t + \varepsilon \tau (x,t,u) + O(\varepsilon ^2 ), \\
\tilde u = u + \varepsilon \phi (x,t,u) + O(\varepsilon ^2 ), \\
\end{array}
\end{displaymath}
where \begin{math}\varepsilon \end{math} is the group parameter.

Use the set of fractional vector fields instead of the one of
integer order
\begin{equation}
V = \xi (x,t,u)D^\beta  _x  + \tau (x,t,u)D^\alpha  _t  + \phi
(x,t,u)D_u, ~~0<\alpha<1,~~0<\beta<1. \label{eq15}
%(15)
\end{equation}

For the fractional second order prolongation \begin{math}Pr^{(2\beta
)} V\end{math} of the infinitesimal generators, we proposed [9]
\begin{equation}
Pr^{(2\beta )} V = V + \phi ^{[t]} \frac{{\partial \phi }}{{\partial
D_t ^\alpha  u}} + \phi ^{[x]} \frac{{\partial \phi }}{{\partial D_x
^\beta  u}} + \phi ^{[tt]} \frac{{\partial \phi }}{{\partial D_t
^{2\alpha } u}} + \phi ^{[xx]} \frac{{\partial \phi }}{{\partial D_x
^{2\beta } u}} + \phi ^{[xt]} \frac{{\partial \phi }}{{\partial D_x
^\beta  D_t ^\alpha  u}}. \label{eq16}
%(16)
\end{equation}

As a result, we can have
\begin{equation}
Pr^{(2\beta )} V(\Delta [u]) = 0, \label{eq17}
%(17)
\end{equation}
 on$\mathop {}\limits^{} \Delta [u]= 0. $

In the time-fractional Burgers equation, Eq. (1), we only need to
consider the case of the fractional order of space $\beta=1$. Thus,
the corresponding Lie algebra of infinitesimal symmetries is the set
of fractional vector fields in the form

\begin{equation}
V = \xi (x,t,u)D_x  + \tau (x,t,u)D_t ^\alpha   + \phi (x,t,u)D_u.
\end{equation}

We assume the one-parameter Lie group of transformations in
\begin{math}{\rm{(}}x,\;t,\;u)\end{math} given by
\begin{equation}
\begin{array}{l}
\tilde{x} = x + \varepsilon \xi (x,t,u)
+ O(\varepsilon ), \\
\frac{{\tilde t^{^\alpha  } }}{{\Gamma (1 + \alpha )}} =
\frac{{t^\alpha  }}{{\Gamma (1 + \alpha )}} + \varepsilon \tau
(x,t,u) + O(\varepsilon ),\\
\tilde u = u + \varepsilon \phi (x,t,u) + O(\varepsilon {\rm{),}} \\
\end{array}
\end{equation}
where \begin{math}\varepsilon \end{math} is the group parameter.

The generalized second prolongation satisfies

\begin{equation} Pr^{(2)} V = V +
\phi ^t \frac{{\partial \phi }}{{\partial D_t ^\alpha
 u}} + \phi ^x \frac{{\partial \phi }}{{\partial D_x u}} + \phi ^{tt}
\frac{{\partial \phi }}{{\partial D_t ^{2\alpha } u}} + \phi ^{xx}
\frac{{\partial \phi }}{{\partial u_{xx} }} + \phi ^{xt}
\frac{{\partial \phi }}{{\partial D_t ^\alpha  u_x }}.
\end{equation}
Using the following condition
\begin{equation}Pr^{(2)} V(\Delta [u]) = 0,~\mathop {}\limits^{} \Delta
[u] = 0,\end{equation} we can have
\begin{equation}
\left. {(\phi ^t  - \phi ^{xx}  - 2u_x \phi ^x )} \right|_{\Delta
[u] = 0}  = 0.
\end{equation}

The generalized prolongation vector fields are reduced as

\begin{equation}
\begin{array}{l}
\phi ^t  = D_t ^\alpha  \phi  - (D_t ^\alpha  \xi )D_x u - (D_t
^\alpha  \tau )D_t ^\alpha  u, \\
\phi ^x  = D_x \phi  - (D_x \xi )D_x u - (D_x \tau )D_t ^\alpha  u,
\\
\phi ^{xx}  = D_x ^2 \phi  - 2(D_x \xi )D_x ^2 u - (D_x ^2 \xi )D_x
u
- 2(D_x \tau )D_x D_t ^\alpha  u - (D_x ^2 \tau )D_t ^\alpha  u . \\
\end{array}
\end{equation}

Substituting Eq. (27) into Eq. (26) and setting the coefficients of
\begin{math}u_{x}u^{(\alpha)}_{xt} ,~u^{(\alpha)}_{xt} \end{math},
\begin{math}u_{xx}u_{x} \end{math}, \begin{math}u_x \end{math} and 1 to
zero. Solve the equations with maple software, we can have

\begin{equation}
\begin{array}{l}
\xi (x,t,u) = c_1  + c_4 x + 2c_5 \frac{{t^\alpha  }}{{\Gamma (1 +
\alpha )}} + 4c_6 \frac{{xt^\alpha  }}{{\Gamma (1 + \alpha )}}, \\
\tau (x,t,u) = c_2  + 2c_4 \frac{{t^\alpha  }}{{\Gamma (1 + \alpha
)}} + 4c_6 \frac{{t^{2\alpha } }}{{\Gamma^{2} (1 + \alpha )}}, \\
\phi (x,t,u) = c_3  - c_5 x + \frac{{2c_6 t^\alpha  }}{{\Gamma (1 +
\alpha )}} - c_6 x^2  + a(x,t)e^u , \\
\end{array}
\end{equation}
where $k^{(\alpha)}_{t}=k_{xx}.$

The Lie algebra of infinitesimal symmetries of Eq. (1) is spanned by
the vector field

\begin{equation}
\begin{array}{l}
V_1  = \frac{\partial }{{\partial x}},\;\;V_2  =
\frac{\partial^{\alpha} }{{\partial t^\alpha  }},\;\;V_3  =
\frac{\partial }{{\partial
u}},\;\;V_4  = x\frac{\partial }{{\partial u}}+\frac{2t^{\alpha}}{\Gamma (1 + \alpha )}\frac{\partial^{\alpha} }{{\partial t^\alpha  }}, \\
V_5  =  - x\frac{\partial }{{\partial u}} + \frac{{2t^\alpha
}}{{\Gamma (1 + \alpha )}}\frac{\partial }{{\partial x}},\;\; \\
V_6  = \frac{{4xt^\alpha  }}{{\Gamma (1 + \alpha )}}\frac{\partial
}{{\partial x}} + \frac{{4t^{2\alpha } }}{{\Gamma^{2} (1 + \alpha
)}}\frac{\partial^{\alpha} }{{\partial t^\alpha  }} - (x^2  +
\frac{{2t^\alpha
}}{{\Gamma (1 + \alpha )}})\frac{\partial }{{\partial u}}, \\
\end{array}
\end{equation}
and the infinite-dimensional subalgebra

$$V_{k}=k(x,t)e^{-u}\frac{\partial}{\partial
u}.$$

It is easy to check the two vector fields \begin{math}\{ V_1 ,V_2
,V_3 ,V_4 ,V_5 ,V_6 \} \end{math} are closed under the Lie bracket
$[a, b]=ab-ba$. In fact, we have

$[V_{i},V_{i}]=0$ $(i=0,...,6)$, $[V_{1},V_{2}]=[V_{1}, V_{3}]=0$,
$[V_{1},V_{4}]=-V_{1}$, $[V_{1},V_{5}]=V_{3}$,

$[V_{1},V_{6}]=-2V_{5}$, $[V_{2},V_{3}]=0$, $[V_{2},V_{4}]=-2V_{2}$,
$[V_{2},V_{5}]=-2V_{1}$, $[V_{2},V_{6}]=2V_{3}-4V_{4}$,

$[V_{3},V_{4}]=[V_{3},V_{5}]=[V_{3},V_{6}]=0$,
$[V_{4},V_{5}]=-V_{5}$, $[V_{4},V_{6}]=-2V_{6}$, $[V_{5},V_{6}]=0.$

$[V_{1},V_{k}]=-V_{k_{x}},$ $[V_{2},V_{k}]=-V_{k_{t}},$
$[V_{3},V_{k}]=-V_{k},$ $[V_{4},V_{k}]=-V_{k^{'}},$
$[V_{5},V_{k}]=-V_{k^{''}},$

$[V_{6},V_{k}]=-V_{k^{'''}},$ \\where
$k^{'}=xk_{x}+\frac{2t^{\alpha}}{\Gamma(1+\alpha)}k^{(\alpha)}_{t},$
$k^{''}=\frac{2t^{\alpha}}{\Gamma(1+\alpha)}k_{x}+xk$ and
$k^{'''}=\frac{4xt^{\alpha}}{\Gamma(1+\alpha)}k_{x}+\frac{4t^{2\alpha}}{\Gamma^2(1+\alpha)}k^{(\alpha)}_{t}+(x^{2}+\frac{2t^{\alpha}}{\Gamma(1+\alpha)})k.$

Take the characteristic equation $V_{5}$ as an example. The
characteristic curve of $V_{5}$ can be given

\begin{equation}
\frac{{du}}{{d\varepsilon}} = -x,
\end{equation}
 \begin{equation}
\frac{dx}{d\varepsilon} = \frac{2t^{\alpha}}{\Gamma (1 + \alpha
)},\label{eq10}
%(10)
\end{equation}
\begin{equation}
\frac{{(dt)^\alpha  }}{{\Gamma (1 + \alpha )d\varepsilon}} = 0. \\
\\
\end{equation}

Solve the above ordinary equations with the initial value
$u=u(x,t,\varepsilon)|_{\varepsilon=0}$,
$x=x(\varepsilon)|_{\varepsilon=0}$ and
$t=t(\varepsilon)|_{\varepsilon=0}$. The one-parameter group $G_{i}$
generated by the $V_{i}$($i= 1,...,6,\alpha$) are given as
\begin{equation}
\begin{array}{l}
g_1 {\rm{:}}\;\;(x,\;\frac{{t^\alpha  }}{{\Gamma (1 + \alpha
)}},\;u) \to (x + \varepsilon ,\;\frac{{t^\alpha  }}{{\Gamma (1 +
\alpha
)}},\;u), \\
g_2 {\rm{:}}\;\;(x,\;\frac{{t^\alpha  }}{{\Gamma (1 + \alpha
)}},\;u) \to (x,\;\frac{{t^\alpha  }}{{\Gamma (1 + \alpha )}} +
\varepsilon
,\;u), \\
g_3 {\rm{:}}\;\;(x,\;\frac{{t^\alpha  }}{{\Gamma (1 + \alpha
)}},\;u) \to  (x,\;\frac{{t^\alpha  }}{{\Gamma (1 + \alpha
)}},\;u+\varepsilon ), \\
g_4 {\rm{:}}\;\;(x,\;\frac{{t^\alpha  }}{{\Gamma (1 + \alpha
)}},\;u) \to (xe^{\varepsilon},\;\frac{{t^\alpha e^{2\varepsilon}
}}{{\Gamma (1 + \alpha )}},\;u),  \\
g_5 {\rm{:}}\;\;(x,\;\frac{{t^\alpha  }}{{\Gamma (1 + \alpha
)}},\;u) \to (x + \frac{{2\varepsilon t^\alpha  }}{{\Gamma (1 +
\alpha )}},\frac{{t^\alpha  }}{{\Gamma (1 + \alpha )}},\;u
-\frac{{\varepsilon ^2 t^\alpha  }}{{\Gamma (1 + \alpha )}} -
x\varepsilon ), \\
g_6 {\rm{:}}\;\;(x,\;\frac{{t^\alpha  }}{{\Gamma (1 + \alpha
)}},\;u) \to (\frac{x }{1-4\varepsilon
\frac{t^{\alpha}}{\Gamma(1+\alpha)}},\;\frac{t^{\alpha}}{1-4\varepsilon
\frac{t^{\alpha}}{\Gamma(1+\alpha)}},\; u-\frac{
x^2\varepsilon}{1-4\varepsilon
\frac{t^{\alpha}}{\Gamma(1+\alpha)}}+\log \sqrt{1-4\varepsilon
\frac{t^{\alpha}}{\Gamma(1+\alpha)}}),\\
g_{\alpha}{\rm{:}}\;\;(x,\;\frac{{t^\alpha  }}{{\Gamma (1 + \alpha
)}},\;u)  \to (x,\;\frac{{t^\alpha  }}{{\Gamma (1 + \alpha
)}},\;\log(e^{u}+\varepsilon k)).
\end{array}
\end{equation}

Take $\alpha=1$ in the above classifications. We can derive the
results for the case of integer order. Since $g_{i}$ is a symmetry,
if $u=f(x, \frac{t^{\alpha}}{\Gamma(1+\alpha)})$ is a solution of
Eq. (1), then the following $u_{i}$ are also the solutions of Eq.
(1)

\begin{equation}
\begin{array}{l}
u_1=f(x -\varepsilon ,\;\frac{{t^\alpha  }}{{\Gamma (1 + \alpha
)}}), \\
u_2=f(x,\;\frac{{t^\alpha  }}{{\Gamma (1 + \alpha )}} - \varepsilon), \\
u_3=f(x,\;\frac{{t^\alpha  }}{{\Gamma (1 + \alpha
)}})+\varepsilon, \\
u_4 =f(xe^{-\varepsilon},\;\frac{{t^\alpha e^{-\varepsilon}
}}{{\Gamma (1 + \alpha )}}), \\
u_5 =f (x - \frac{{2\varepsilon t^\alpha  }}{{\Gamma (1 + \alpha
)}},\frac{t^\alpha }{\Gamma (1 + \alpha)}) + \frac{{\varepsilon ^2
t^\alpha  }}{{\Gamma (1 + \alpha )}} -
x\varepsilon, \\
u_6=f(\frac{x }{1+4\varepsilon
\frac{t^{\alpha}}{\Gamma(1+\alpha)}},\;\frac{t^{\alpha}}{1+4\varepsilon
\frac{t^{\alpha}}{\Gamma(1+\alpha)}}\;)
-\frac{x^2\varepsilon}{1-4\varepsilon
\frac{t^{\alpha}}{\Gamma(1+\alpha)}}-\log \sqrt{1-4\varepsilon
\frac{t^{\alpha}}{\Gamma(1+\alpha)}},\\
u_{\alpha}=\log(e^{f(x,\frac{t^{\alpha}}{\Gamma(1+\alpha)})}+\varepsilon
k).
\end{array}
\end{equation}

Now we consider the applications of the above transformations. From
$u_{1}$ to $u_{4}$, we can only obtain trivial solutions. Therefore,
we start from the use of $u_{5}$

\begin{equation}g_{5}{\rm{:}}\;\;(x,\;\frac{{t^\alpha }}{{\Gamma (1
+ \alpha )}},\;u) \to (x + \frac{{2\varepsilon t^\alpha }}{{\Gamma
(1 + \alpha )}},\frac{{t^\alpha  }}{{\Gamma (1 + \alpha )}},\;u -
\frac{{\varepsilon ^2 t^\alpha }}{{\Gamma (1 + \alpha )}} -
x\varepsilon ),\end{equation}

Assume \begin{math}u_{5,0}=u_{5,0}(x,\;\frac{{t^\alpha  }}{{\Gamma
(1 + \alpha )}}) = f(x,\;\frac{{t^\alpha  }}{{\Gamma (1 + \alpha
)}})\end{math} is one exact solution of Eq. (1). Take $u_{5,0}=c$,
where $c$ is a arbitrary constant and also a trivial solution.  We
can get a new nontrivial exact solution as

\begin{equation}u_{5,1}=c + \frac{{\varepsilon ^2
t^\alpha }}{{\Gamma (1 + \alpha )}} - x\varepsilon .\end{equation}

Further more, continue this iteration process, we can derive a new
exact solution of Eq. (1)

\begin{equation}u_{5,2}=u_{5,1}(x - \frac{{2\varepsilon
t^\alpha }}{{\Gamma (1 + \alpha )}},~\frac{{t^\alpha }}{{\Gamma (1 +
\alpha )}}) + \frac{{\varepsilon ^2 t^\alpha }}{{\Gamma (1 + \alpha
)}} - x\varepsilon
=c-2x\varepsilon+\frac{4\varepsilon^{2}t^{\alpha}}{\Gamma (1 +
\alpha )}.\end{equation}

Similarly, take $u_{6,0}=c$, then we can have
\begin{equation}u_{6,1}=c -\frac{x^2\varepsilon}{1+4\varepsilon
\frac{t^{\alpha}}{\Gamma(1+\alpha)}}-\log \sqrt{1+4\varepsilon
\frac{t^{\alpha}}{\Gamma(1+\alpha)}},\end{equation}

and
\begin{equation}u_{6,2}=u_{6,1}(\frac{x }{1+4\varepsilon
\frac{t^{\alpha}}{\Gamma(1+\alpha)}},\;\frac{t^{\alpha}}{1+4\varepsilon
\frac{t^{\alpha}}{\Gamma(1+\alpha)}}\;)
-\frac{x^2\varepsilon}{1+4\varepsilon
\frac{t^{\alpha}}{\Gamma(1+\alpha)}}-\log \sqrt{1+4\varepsilon
\frac{t^{\alpha}}{\Gamma(1+\alpha)}}\end{equation}

~~~~~~~~\noindent{$\displaystyle  =c-\frac{2\varepsilon x^{2}}{1+
\frac{8\varepsilon t^{\alpha}}{\Gamma(1+\alpha)}}-\log\sqrt{1+
\frac{8\varepsilon t^{\alpha}}{\Gamma(1+\alpha)}}. $}

 We can readily verify that
$u_{5,1}$, $u_{5, 2}$, $u_{6,1}$ and $u_{6, 2}$ are four exact
solutions of Eq. (1). Assuming the fractional order $\alpha=1$, the
exact solutions we give here can be reduced as the exact iteration
solution in Ref. [16] if we set the coefficients $a=b=1$.

\section{Conclusions}

Fractional differential equations have caught considerable attention
due to their various applications in real physical problems.
However, there is no a systematic method to derive the exact
solutions. Now, the problem is partly solved in this paper. The
presented paper can be applied to other fractional partial
differential equations and investigate their non-smooth properties.

%--------------------------------------------------
\vskip 10 pt \leftline{\bf\ Acknowledgments}

The author would like to give his thanks to Prof. En-gui Fan (Math.
Dep., Fudan University, China) and the referee's sincere
suggestions. The author also feels grateful to Dr. Hong-li An's
helpful discussions when visiting Hongkong Polytechnic University
this summer.

%\end{CJK}
\clearpage

\end{document}